# Photoexcitation-induced Stacking Transition Assisted by Intralayer Reconstruction in Charge-Density-Wave Materials

Jin Zhang[1,*], Yang Yang[2], Jia Zhang[3], Mengxue Guan[4], Jiyu Xu[2,5], Kun Yang[1], Xinghua Shi[1,*], Sheng Meng[2,5,*]

[1]*Laboratory of Theoretical and Computational Nanoscience, National Center for Nanoscience and Technology, Chinese Academy of Sciences, Beijing 100190, P. R. China*
[2]*Beijing National Laboratory for Condensed Matter Physics, and Institute of Physics, Chinese Academy of Sciences, Beijing 100190, P. R. China*
[3]*Max Born Institut für Nichtlineare Optik und Kurzzeitspektroskopie, Berlin 12489, Germany*
[4]*Centre for Quantum Physics, Key Laboratory of Advanced Optoelectronic Quantum Architecture and Measurement (MOE), School of Physics, Beijing Institute of Technology, Beijing 100190, P. R. China*
[5]*Songshan Lake Materials Laboratory, Dongguan, Guangdong 523808, China*

\* Corresponding authors: Jin Zhang (jinzhang@nanoctr.cn),
Xinghua Shi (shixh@nanoctr.cn),
Sheng Meng (smeng@iphy.ac.cn)

Laser excitation has emerged as an effective tool for probing microscopic interactions and manipulating phases of matter. Among charge density wave (CDW) materials, 1$T$-TaS$_2$ has garnered significant attention due to its diverse stacking orders and photoexcited responses. However, the mechanisms driving transitions among different stacking orders and the microscopic out-of-equilibrium dynamics remain unclear. We elucidate that photoexcitation can introduce interlayer stacking order transitions facilitated by laser-induced intralayer reconstruction in 1$T$-TaS$_2$. Importantly, our finding reveals a novel pathway to introduce different phases through laser excitations, apparently distinct from thermally-induced phase transitions via interlayer sliding. In particular, photoexcitation is able to considerably change potential energy surfaces and evoke collective lattice dynamics. Consequently, the laser-induced intralayer reconstruction plays a crucial role in interlayer stacking-order transition, offering a new method to create exotic stackings and quantum phases. The exploration opens up great opportunities for manipulating CDW phases and electronic properties on the femtosecond timescale.

Interactions among different degrees of freedom, including electrons, phonons, and other quasiparticles, are pivotal in understanding and manipulating the properties of quantum materials [1-3]. Optical driving with strong laser pulses facilitates the identification of predominant interactions and provides valuable insights into ground-state properties, phase transitions, and hidden states [4-10]. The ability to modulate properties of charge density wave (CDW) materials using light holds great potential for its future applications in information technology [11-15]. Notably, 1$T$-TaS$_2$ stands out as an intensively investigated prototypical CDW material for its complex electronic phase diagram, which includes temperature-dependent metal-to-insulator transition as well as the emergence of pressure-induced superconductivity [16-21].

Recently, substantial efforts have been made to elucidate the microscopic mechanism of phase transitions in 1$T$-TaS$_2$, focusing on stacking orders and non-equilibrium insulator-to-metal transitions [5, 22-28]. Utilizing ultrashort laser pulses, the CDW state can be suppressed and the closure of the band gap takes place within tens of femtoseconds in 1$T$-TaS$_2$ [5,14]. Interestingly, Ritschel *et al.* revealed metastable interlayer stackings and intriguing orbital textures, which significantly modulate the electronic structures [22]. Theoretically, it is observed that charge density wave reconstruction on the surface, involving small atomic displacements in the subsurface layer, apparently modifies the surface electronic structure [23]. In addition, Vaňo *et al.* demonstrated that Kondo coupling between localized magnetic moments and itinerant electrons creates artificial heavy fermions in the TaS$_2$ heterostructure, revealing a hybridization gap or a doped Mott insulator [24, 25]. However, the microscopic coupling among stacking orders, ionic movements, and electronic modulations remains enigmatic.

In this Letter, we investigate photoexcited stacking transition in the prototypical charge-density-wave material 1$T$-TaS$_2$, employing time-dependent density functional theory molecular dynamics (TDDFT) simulations [29]. The calculations allow us to unravel photoinduced dynamics on both the atomic scale and femtosecond timescale. Through optical excitations, the potential surface is significantly modulated and collective lattice dynamics of CDW are invoked with different patterns. Additionally, we identify intrinsic processes of atomic rearrangements of Star-of-David (SoD) patterns and diverse interlayer stackings are attainable from laser photoexcitation. Specifically, stacking orders can be engineered by controlling the strengths and frequencies of laser pulses applied. The study not only offers profound insights into ultrafast phase transitions of CDW materials but also lays the ground for a new framework to comprehend photoinduced phenomena in a broad array of quantum materials.

1$T$-TaS$_2$ is a quasi-two-dimensional CDW material, whose origin is commonly attributed to Fermi-surface nesting driven by electron-phonon coupling through the Peierls mechanism [4-6]. As the temperature drops below 350 K, SoD patterns of tantalum (Ta) atoms emerge, accompanied by a periodic distortion with a √13×√13 superlattice. The reconstruction brings about an insulating electronic structure due to the presence of a fully commensurate CDW structure. In the system, Ta atoms and the surrounding S atoms move inward synchronously, resulting in charge-density accumulation and the appearance of charge centers.

In the CDW phase of 1$T$-TaS$_2$, three types of Ta atoms are identified in the supercell [indicated in Fig. (1)]: one at the center of the SoD pattern (labeled as Ⓐ sites in panel c), six at the adjacent rings (labeled as Ⓑ sites), and the other six Ta atoms at the subsequent outer rings (labeled as Ⓒ sites). Charge transfer from the outer-ring Ta atoms (Ⓒ or Ⓑ sites) towards the center (Ⓐ sites) results in apparent lattice distortion and a larger supercell. The interlayer stacking orders are dependent on the relative arrangements of Ta atoms. For clarity, we first consider simple cases assuming two layers in a unit cell although the real stackings can be very complex in the CDW phase of 1$T$-TaS$_2$ [22-23]. From Figs. 1(a-d) we present the atomic structures of various 1$T$-TaS$_2$ phases. Based on the arrangement of the SoD patterns, the CDW phases are categorized into AA, AB, AC, AACC, and other configurations. In the AA stacking, the centers of the SoD patterns are aligned directly on top of each other across adjacent layers, which is the most commonly observed phase in experiments [6,19]. Between AA and AB stackings, there is an interlayer shift of approximately 3.15 Å between the adjacent layers, whereas the Ta atoms experience an interlayer displacement of around 5.57 Å from AA to AC stacking. Large atomic displacements naturally mean high energy barriers for the interlayer sliding processes as schematically shown in Fig. 1(e). In our calculations, the energy barriers are obtained with the nudged elastic band optimization by interpolating structures to find the saddle point of the transition path.

In comparison, CDW phases with various stacking orders can readily transform into unmodulated 1$T$ phase via thermal excitation or photoinduced dynamics (*i.e.*, melting of CDW states). In this regard, we propose an alternative strategy to facilitate phase transitions among stacking orders: intralayer reconstruction of SoD patterns [Fig. 1(f)]. Firstly, the CDW structure converts into the undistorted phase with a 1×1 periodic lattice, causing the disappearance of the charge centers. Subsequently, the Ta atoms or charges nucleate and generate new SoD centers.

The stabilization of SoD centers and the CDW transition are paramount to the success of the strategic approach.

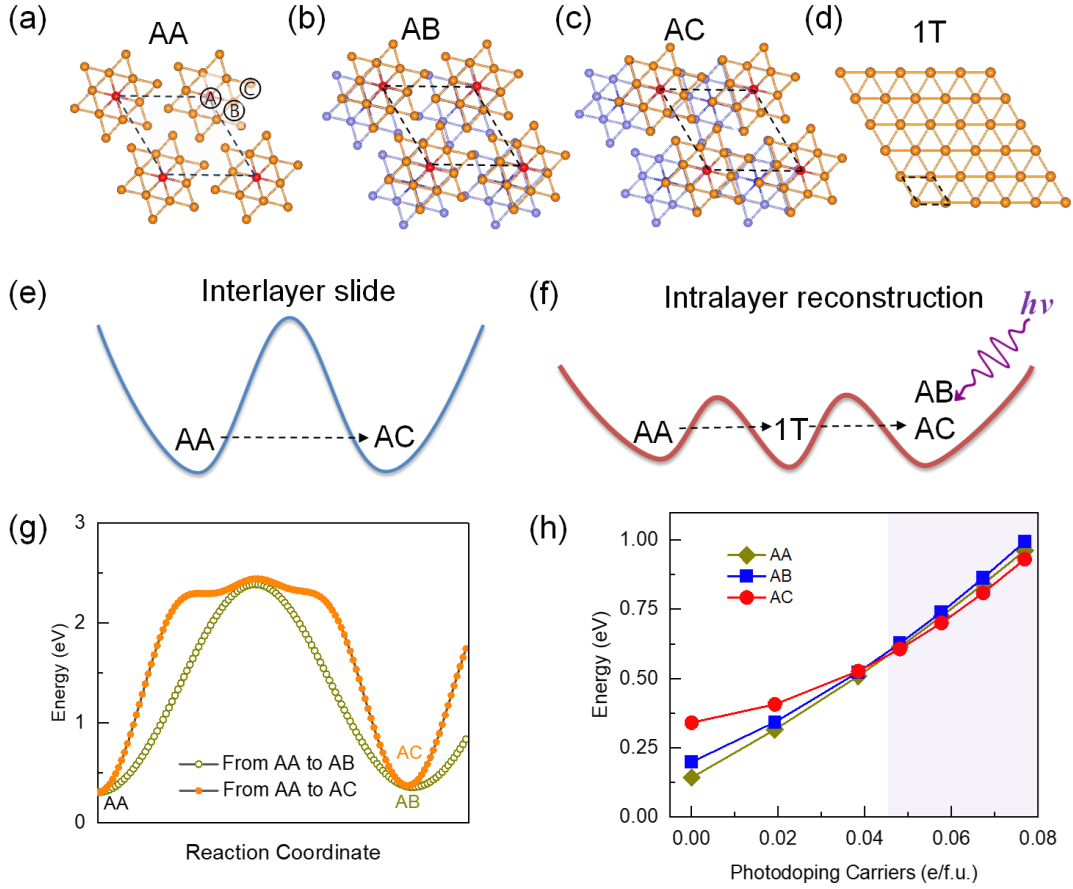

**FIG. 1. Structures and potential surfaces of 1$T$-TaS$_2$ with various stacking orders.** (a-c) Atomic structure of 1$T$-TaS$_2$ in AA, AB and AC stacking orders with a √13 × √13 superlattices. Three sites of Ta atoms are labeled with Ⓐ, Ⓑ and Ⓒ, respectively. (d) Atomic structure of 1T pristine phase without periodic reconstruction. Orange and light blue spheres denote Ta atoms, while red and blue spheres are Ta atoms at the center of stars, respectively. The sulfur atoms are not shown for clarity. The criterion for making bonds is that the Ta-Ta distance is shorter than 3.4 Å. The unit cells are illustrated with dashed rhombuses in panels (a-d). (e) Schematic potential surface for the thermal transition and interlayer slide among different stacking orders. (f) The pathway for photoexcited phase transition via "intralayer reconstruction" mechanism. (g) Energy surfaces for interlayer slides from AA to AB and AC stackings, respectively. (h) Energetic profile for the three stacking orders under different photodoping levels.

At the ground state, the CDW phase exhibiting AA stacking order proves to be the most energetically favorable, with energy levels of AB and AC stackings being 56 and 98 meV higher, respectively. The profiles of the potential surfaces are calculated to facilitate a comparative analysis of the two pathways, as shown in Fig. 1(g). The energy barrier for interlayer sliding from AA to AC stacking is 2.13 eV, while the transition from AA to AB stacking encounters a slightly lower barrier of 2.08 eV. Realizing the transitions via thermal methods is remarkably challenging due to the substantial barriers. The thermal stability of the CDW phases is further scrutinized for different photodoping levels, wherein photodoped carriers are artificially excited from states below the Fermi level to those above, utilizing constrained DFT simulations with fixed occupations. At a photodoping level of 0.048 e/f.u. (*i.e.*, electrons per formula unit or electrons per Ta atom), the total energies for the three configurations converged, indicating a comparable degree of stability. At elevated doping levels, the thermal stability associated with AC stacking surpasses that of AA stacking, implying the emergence of a new configuration, as illustrated in Fig. 1(h).

We further explore the time-dependent structural dynamics and phase transition of CDW phases under different photoexcited environments. Fig. 2(a) shows the applied laser pulses with Gaussian-envelope function and the photon energy of 1.0 eV, which is in the range of widely adopted optical measurements. We meticulously prepare the initial structures of 1*T*-TaS$_2$ by introducing random structural distortions to break structural symmetry, representing the intermediate states of the proposed mechanisms. In the pristine 1*T* phase, all Ta atoms are equivalent, establishing a local energy minimum on the potential surface. To break the structural symmetry, specific Ta atoms are designated as central, while others are classified as peripheral with extended Ta-Ta distances. The amplitude of the random structural distortions approximates 0.1 Å, notably smaller than the distortions obtained during the CDW transition.

Previous studies have identified that the critical temperature for melting the CDW phase is at approximately 400 K using Born-Oppenheim molecular dynamics [7]. Beyond this threshold, the typical star-shaped lattice distortions vanish, reverting to the undistorted 1*T* geometry where spatial modulation weakens significantly. Fig. S1 in the Supplementary Materials displays snapshots taken at the ionic temperatures of 100 K and 300 K from Born-Oppenheimer molecular dynamics. After the evolution of 1 picosecond, the structures transform into disordered states without long-range periodicity at both temperatures. The ab initio molecular dynamics (MD) simulations are performed to examine the thermal stability under different ionic temperatures, validating that the initial displacements are not stable within finite timescales. This suggests that thermal excitations at these temperatures can activate multiple phonons, thereby suppressing the ordered CDW transition within an ultrafast timescale.

From Fig. 2(b) we determine the number of carriers excited from the valence to the conduction bands upon optical illumination. In the beginning, the number of carriers fluctuates with the laser pulses. Then it stabilizes after 20 fs as the laser fields diminish. For an electric field of $E_0$=0.013 V/Å, the number of photoexcited carriers reached 0.022 e/f.u. As the intensity increases, the number of carriers rises obviously. When $E_0$ reaches 0.05 V/Å, the carrier density is as high as 0.269 e/f.u. See the supplementary materials (Fig. S2) for the distribution of the excited carriers. This observation implies that laser excitation can effectively stimulate variable densities of electrons and holes, thereby altering the underlying potential landscapes. As shown in Fig. S3, we observe the energy barriers for transformation drop to 0.8 and 0.2 eV from AA to AB and AC stacking, respectively. In addition, the atomic displacement for the stacking rearrangement is only about 0.3 Å and the energy barrier is much smaller (several meV) under certain photoexcitation (*e.g.*, $E_0$=0.05 V/Å).

Subsequently, the photoinduced dynamics in 1*T*-TaS$_2$ are investigated across varying laser intensities and frequencies. To elucidate the structural transformations, the root mean square displacements (RMSD) are analyzed [see Fig. 2(c)], which is a metric that quantifies the extent of structural changes. Notable modulations in atomic positions are noticed, with the RMSD increasing to 0.4 Å after approximately 420 fs under the low intensity ($E_0$=0.013 V/Å) with a photon energy of 1.0 eV. The pronounced structural changes at low fluences indicate the thermal instability of initial structures with initial distortions. For a higher intensity of $E_0$=0.05 V/Å, the electron-nuclear dynamics exhibits distinct features, in which the RMSD decreases to 0.32 Å, along with periodic oscillations. The lattice dynamics are attributed to the amplitude mode of the CDW (the period of ~400 fs), where atomic clusters and charge order synchronously vibrate around their equilibrium positions [Fig. S4]. We note that the modulation of RMSD is in good agreement with the experimental observation [5]. The electronic and structural dynamics can be detected and tracked by time-resolved angle-resolved photoemission spectroscopy and femtosecond *X*-ray measurement [5,8]. This finding highlights that photoexcitation can initiate collective lattice dynamics and promote spatially ordered atomic patterns in 1*T*-TaS$_2$.

In addition, Fig. 2(d) shows the effective ionic temperatures under laser excitations. The oscillations are interpreted as the coherent exchange of the kinetic and ionic potential energy. Our TDDFT-MD simulations naturally account for the damping of high-energy quasiparticles. These quasiparticles dissipate their energy to electrons at lower energy levels and to the ionic subsystem [7]. For the intensities of $E_0$ = 0.013 and 0.025 V/Å, the lattice temperatures rise to 250 K, during which excited carriers transfer their energy to the lattice system through effective electron-phonon interactions. Nevertheless, no SoD patterns or CDW transitions are obtained, as illustrated in Fig. 2(f). In regard to the intensity of $E_0$=0.05 V/Å, we observe an effective temperature decrease occurring after 200 fs. In the condition, well-defined SoD patterns are found, indicating that the excitation energy is sufficient to provoke a phase transition from an initial distorted structure to a long-range CDW configuration, as depicted in Fig. 2. Besides, the frequency-dependent structural dynamics are investigated (Fig. S5). Different photon energies with energies of 0.5 eV and 1.0 eV are compared to confirm the findings discussed above.

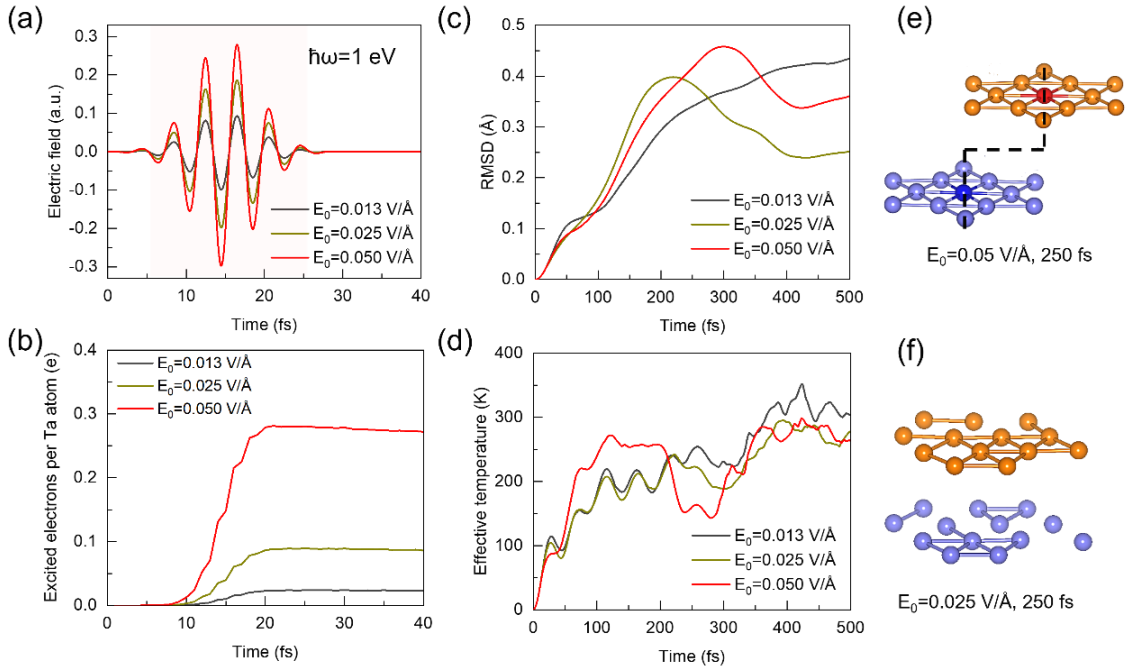

**FIG. 2. Laser-induced excited dynamics in 1*T*-TaS$_2$.** (a) Applied electric fields along the in-plane direction. (b) The number of excited electrons upon photoexcitation from the valence bands to conduction bands under the laser pulses. (c) Root mean square displacement (RMSD) of 1*T*-TaS$_2$ from TDDFT simulations. (d) Time evolution of lattice temperatures. (e) Snapshots for laser-induced state for the laser strength $E_0$= 0.05 V/Å at 250 fs. (f) The same quantity as (e) for the laser strength $E_0$=0.025 V/Å at 250 fs, the CDW order is not introduced for the intensity.

To gain more insights into the structural dynamics triggered by photoexcitation, Fig. 3(a) exhibits the snapshots of structural changes at different times after optical illumination for $E_0$= 0.05 V/Å. At t=0 fs, the pristine phase with small atomic displacements of TaS$_2$ is excited. The main peak at 3.36 Å of the radical distance distributions corresponds to the signature value of the undistorted phase. Other peaks are attributed to the initial atomic displacements. With time evolution, we observe that the Ta-Ta radical distributions change apparently and the broadening becomes narrower at 150 fs. Afterward, the typical Ta-Ta distributions appear at 3.20 Å, coming from the Ta-Ta distances between center Ta (Ⓐ sites) and the inner rings (Ⓑ sites). Another important peak is around 3.90 Å, which is attributed to the Ta-Ta distances among Ⓑ and Ⓒ sites, as shown in Fig. 3(b). For comparison, the distribution of Ta-Ta distances is characterized by the peaks at 3.20 Å and 3.90 Å in the pristine CDW state. If the number of excited carriers is enough to overcome the disordered transition, a spontaneous reconstruction of the CDW states is triggered, causing a different stacking order in one layer to relocate to the other layer.

In addition to the photoinduced collective atomic motions in 1$T$-TaS$_2$, valuable information is also gained from the evolution of electron diffraction patterns during the coupled electron-lattice evolution caused by photoexcitation. Fig. 3(c) exhibits simulated diffraction patterns for the transient states, with typical patterns for the undistorted structure shown (blue circles) and those for the laser-induced CDW patterns indicated in red circles. Our spatial structural snapshots, together with the Ta-Ta radical distributions confirm that the collective lattice dynamics of the CDW structures with different stackings are invoked by photoexcitation [8].

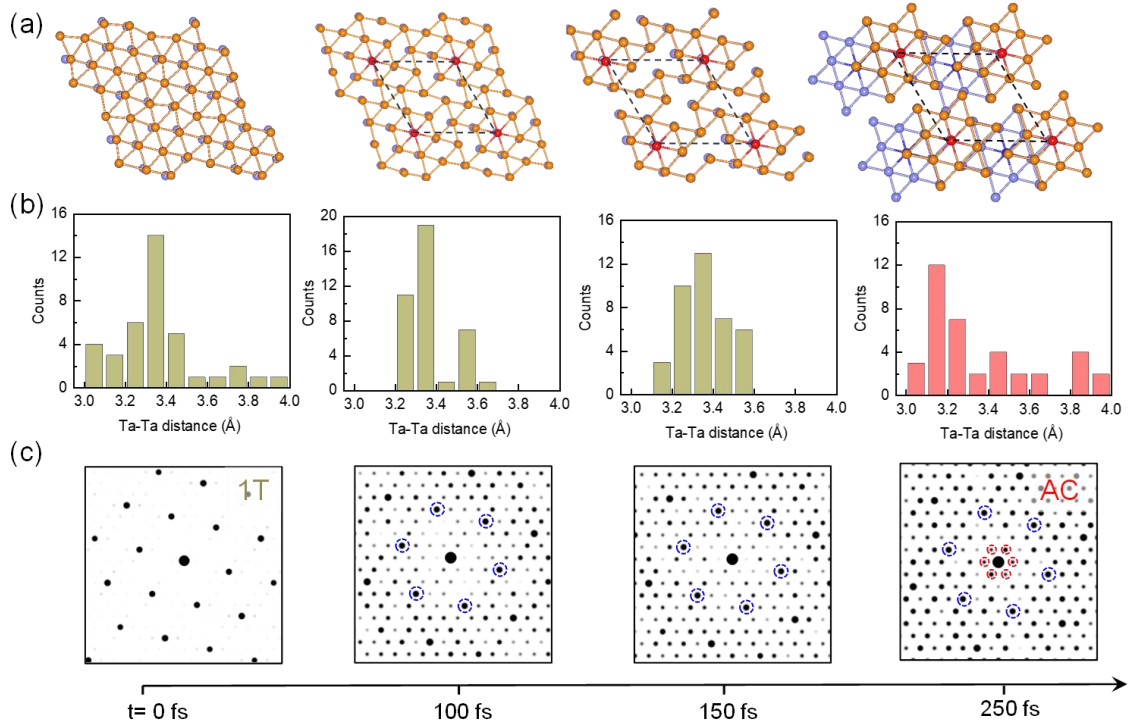

**FIG. 3. Time evolution of structures of 1$T$-TaS$_2$ under photoexcitation.** (a) Snapshots for time-dependent atomic structures after photoexcitation, respectively. The laser strength is $E_0$= 0.05 V/Å. (b) Corresponding radial distribution of Ta-Ta distances. (c) Simulated electron diffraction patterns for the transient states, with typical patterns for the undistorted structure shown in blue and those for the laser-induced CDW patterns shown in red.

For the intensity of $E_0$= 0.05 V/Å, the photoexcited dynamics exhibit distinct trends, despite the lattice temperature higher than 200 K. A photoinduced transient state emerges at 210 fs with an RMSD of 0.35 Å. Interestingly, increasing the intensity does not significantly influence the collective dynamics. Clear CDW orders are found for laser intensities exceeding 0.05 V/Å. After 250 fs, 1$T$-TaS$_2$ experiences obvious structural modulation, resulting in a lattice distortion with a larger periodicity of approximately 12.1 Å, rotated by 13.9° compared with the 1$T$ phase. Particularly, the laser pulse triggers collective lattice dynamics, entailing synchronous oscillations of atomic clusters and charge order around their equilibrium positions. Besides, strong laser pulses can bring about a variety of CDW phases in 1$T$-TaS$_2$, which are characterized by periodic modulations in the electronic charge density and coupled to distortions in the lattice. The optically induced structural changes arise from electron-phonon coupling following optical excitation, as illustrated in Fig. S6.

Furthermore, we compare the electronic properties of transient structures during the real-time propagation to dig out more information about the photoexcited states. Fig. S7 illustrates the evolution of band structures at various times after photoexcitation, derived from snapshots of photoexcited 1$T$-TaS$_2$ under the strength of E$_0$ = 0.05 V/Å. The band structures unveil an in-plane gap accompanied by interlayer metallic behavior, suggesting pronounced modulation in both atomic structure and electronic properties.

In the context of initial distortions, thermal excitations are able to cause fluctuations in atomic positions, accounting for the formation of the charge accumulation. In addition, the accumulation and depletion of charge densities significantly impact atomic displacements [Figs. 4(a-c)]. This approach involves interfacial carrier doping, which can be achieved by adsorbing some molecules (*e.g.*, H$_2$O) or by intercalating lithium ions [37,38], which effectively change the charge distribution at interfaces, thereby influencing atomic displacements and resulting in the new centers of SoD patterns.

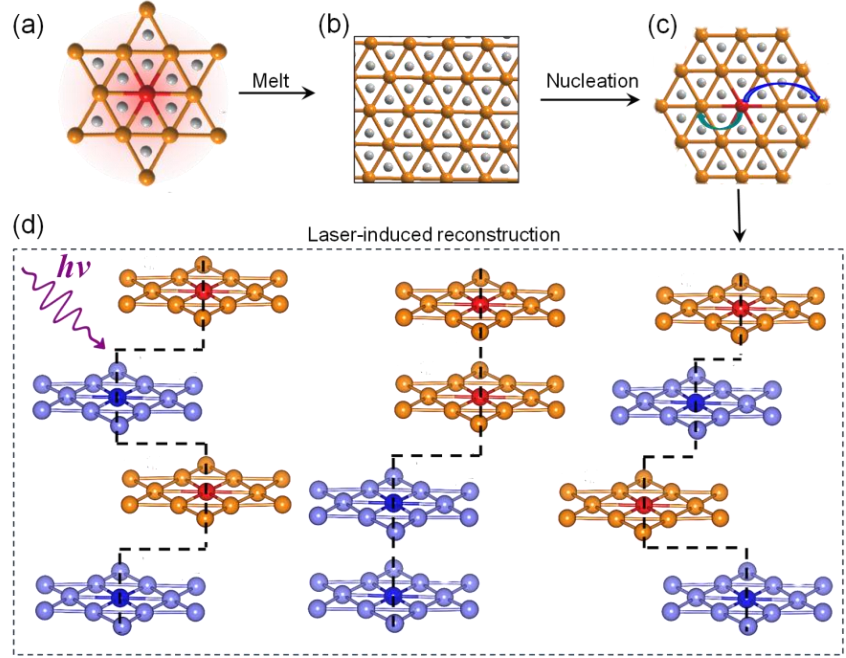

**FIG. 4. Photoexcitation-induced stacking transition facilitated by intralayer reconstruction in 1$T$-TaS$_2$.** (a) The melting of the CDW state. (b) The undistorted 1$T$ state. Orange and grey spheres represent Ta and S atoms, respectively. The Ta atoms at the centers of SoD patterns are highlighted in red. (c) The nucleation of SoD patterns at different sites. The arrows in the panel denote the reconstruction of SoD patterns. (d) The laser-induced phase transition with the AC, AL, and an arbitrary stacking order (ABCA) in 1$T$-TaS$_2$, respectively. Dashed black lines serve as guides for the interlayer stacking

orders.

The exploration of photoinduced dynamics extends to additional stackings by incorporating four layers with varying Ta displacements. Notably, under the identical laser pulse conditions ($E_0 = 0.05$ V/Å and a photon energy of 1 eV), a specific bilayer CDW configuration, known as AL stacking (AACC), as depicted in Fig. 4(d), is recognized as the most stable phase and validated by numerous studies [41,42]. The modulation in vertical CDW alignment for every two layers effectively generates Peierls dimerization in the vertical chains of the half-filled localized states, rendering an insulating state [41]. Moreover, we obtain an arbitrary stacking order (namely, ABCA) by incorporating various stacking of Ta centers. The structures in Figure 4(d) are precisely obtained from the TDDFT simulations, which is elaborated in Fig. S8. For the statistical analysis, additional calculations with different initial distortions, reveal that an initial atomic displacement larger than 0.1 Å is required. Otherwise, the initial distortions would be destroyed by the elevated temperature and the nucleation takes much longer time which is too expensive and beyond the capability of TDDFT simulations. The findings underscore the robustness of laser-induced phase reconstruction in CDW materials with more complicated layer freedom, transcending variations in configurations, highlighting the potential of employing short laser pulses as a precise and controllable tool for manipulating the phases in CDW materials.

In addition, a continuous transition from AA stacking to 1T phase and AC stacking is very helpful to the physical picture. However, it is nearly not possible in real-time simulations, as shown in Fig. S9. We performed additional calculations with larger supercells, validating our main findings [Fig. S10]. Notably, our first-principles simulations are not able to describe the experimentally observed incommensurate structures and topological defects on longer timescales following photoexcitation [4,20]. They are beyond the primary focus of our study because these processes typically occur on much longer timescales and larger spatial scales. In CDW materials, a phason is a low-energy collective mode associated with displacements in the periodic charge density waves. It is a quasiparticle associated with the collective oscillations or shifts in the phase of a periodic lattice [43]. The mechanism of the laser-induced phase transition can be understood as photodoping-facilitated dynamics of phasons. Consequently, our results provide crucial support and complementary insights to fully grasp the atomic processes in the ultrafast phase transition.

Therefore, we present a comprehensive picture of how laser-induced intralayer reconstruction facilitates stacking transitions in CDW materials. Following thermal or optical melting of the CDW, various SoD centers nucleate, assisted by initial random distortions. This process requires an initial atomic displacement exceeding 0.1 Å. The nucleation occurs in the 1T phase with arbitrary centers, breaking the geometry symmetry in the spatial distribution of Ta atoms. Strong laser pulses can introduce numerous photodoping carriers into undistorted $TaS_2$ phases, thereby modulating potential surfaces and leading to varied lattice instability. Laser-induced reconstruction is realized by photodoped carriers with a density exceeding 0.25 e/f.u. Consequently, SoD centers undergo significant modulations, giving rise to distinct configurations and electronic structures. The results clarify the microscopic mechanism of the photoinduced phase transition of CDW materials and shed light on a broader spectrum of quantum materials.

Our *ab initio* TDDFT simulations demonstrate the photoinduced stacking-order phase transitions through intralayer atomic reorganization in 1*T*-$TaS_2$. These findings show that ultrafast reconstruction of CDW phases can be realized by short laser pulses, with photoexcited carriers playing a crucial role. Moreover, we identify SoD rearrangements and different configurations that are attainable through intralayer reconstruction under photoexcitation. The study offers a new understanding of laser-triggered stacking reorganization and phase transitions in layered CDW materials, advancing our current understanding of correlated materials and opening new avenues for their application in next-generation electronic devices.

**Acknowledgements**

This work was supported by the National Key R&D Program of China (2022YFA1203200), the Basic Science Center Project of the National Natural Science Foundation of China (22388101), the Strategic Priority Research Program of the Chinese Academy of Sciences (XDB36000000), the National Natural Science Foundation of China (12125202).